\documentclass[useAMS,usenatbib]{mn2e}
\usepackage{psfig, epsf, epsfig, aas_macros}
\def\slantfrac#1#2{\kern.1em^{#1}\kern-.3em/\kern-.1em_{#2}}

\title[Heart of Darkness]{Quantifying the heart of darkness with GHALO - a multi-billion particle simulation of our galactic halo}
\author[Joachim Stadel et al.]{
J.~Stadel$^1$,
D.~Potter$^1$,
B.~Moore$^1$,
J.~Diemand$^2$,
P.~Madau$^2$,
M.~Zemp$^2$,
\newauthor
M.~Kuhlen$^3$ \&
V.~Quilis$^4$\\
       ${}^1$Institute for Theoretical Physics, University of Zurich,
             Winterthurerstr. 190, 8057 Zurich, Switzerland\\
       ${}^2$University of California, Department of Astronomy and Astrophysics, 1156 High Street, Santa Cruz CA 95064, USA\\
       ${}^3$Institute for Advanced Study, Einstein Drive, Princeton, NJ 08540, USA\\
       ${}^4$Departament   d'Astronomia  i  Astrof\'{\i}sica,
       Universitat  de   Val\`encia,  Dr.~Moliner  50,   46100  Burjassot (Val\`encia),  Spain
    }

\begin{document}

\date{Accepted, Received; in original form }

\pagerange{\pageref{firstpage}--\pageref{lastpage}} \pubyear{2008}

\maketitle

\label{firstpage}

\begin{abstract}
We perform a series of simulations of a Galactic mass dark matter 
halo at different resolutions, our largest uses over three billion particles and has a mass resolution 
of $1000 M_\odot$. We quantify the structural properties of the inner dark matter distribution and study how they depend on numerical resolution. 
We can measure the density profile to a distance of 120 pc (0.05\% of $R_{\rm vir}$) where the logarithmic slope is -0.8 and -1.4 at (0.5\% of $R_{\rm vir}$). 
We propose a new two parameter fitting function that has a linearly varying logarithmic density gradient which fits the GHALO and VL2 density profiles extremely well.
Convergence in the density profile and the halo shape scales as $N^{-1/3}$, but the shape converges at a radius three times larger at which point the halo becomes more spherical due to numerical resolution. 
The six dimensional phase-space profile is dominated by the presence of the substructures and 
does not follow a power law, except in the smooth under-resolved inner few kpc. 
%A network of overlapping streams, dark debris from orbiting satellites can be observed filling phase-space.

\end{abstract}


\section{Introduction}

Over twenty five years ago the theoretical framework for the evolution of a cold dark matter 
(CDM) dominated universe was established \citep{1982ApJ...263L...1P}. The hierarchical and violent 
growth of structure in this model begins at a scale of $~10^{-6} M_\odot$ 
\citep{2005Natur.433..389D} 
until the most massive clusters of galaxies form that are many orders of 
magnitude more massive. The assumption that the dark matter is cold remains to be verified, 
yet numerical simulations that follow the hierarchical formation of CDM haloes have given 
several fundamental and robust predictions for the structural and substructure properties 
of the dark matter distribution within virialised haloes 
\citep{1991ApJ...378..496D,1996ApJ...462..563N,1999MNRAS.310..527A,2001MNRAS.321..559B,2001ApJ...557..533F}. 
These results are widely used 
to compare with observational data and to assist comparisons with analytic models.
%, and over the past 15 years to critically test the CDM paradigm.

%Over the past decade our standard cosmological model has suffered one if its most serious 
%challenges when critically compared to observations of galactic halo structure and 
%substructure. Most of the current interest has been in the global density profiles and 
%the mass function and kinematics of dark matter substructures that can be associated and 
%compared with satellite galaxies. Whether or not CDM haloes provide a good description of 
%the dark matter surrounding galaxies has been debated since 1994.  
%However, recent 
%observations from deep sky surveys have uncovered a host of barely visible yet dark matter 
%dominated substructures within the inner regions of the Milky Way.

The first CDM halo simulated with enough resolution to resolve 
substructure used $10^6$ particles \citep{1998ApJ...499L...5M}, 
resolving the density profile to about one percent of 
the virial radius \citep{1998MNRAS.300..146G,1999MNRAS.310.1147M,2003MNRAS.338...14P,2004MNRAS.349.1039N,
2004MNRAS.353..624D}.
%, equivalent to 3 kpc in the 
%Galaxy or $\sim 500$ pc within the Fornax dSph.
Whilst such simulations find numerous substructures in the outer halo, they find few or none 
within the inner 20\% of $R_{\rm vir}$ and no obvious structure in phase-space in the central halo 
regions \citep{2001PhRvD..64f3508M}. 
Advances in algorithms and supercomputing power have recently allowed us to increase this resolution by over 
two orders of magnitude with the Via Lactea II (VL2) simulation \citep{2008arXiv0805.1244D}.
%These authors followed a billion particles in total with a convergence radius of $\sim 400$ parsecs (0.1\% of $R_{\rm vir}$). 
%Once high enough force and integration accuracy is achieved, 
%The convergence radii of the density profile scales as $N_{\rm vir}^{-1/3}$, where $N_{\rm vir}$ is the number 
%of particles in the virial radius \citep{2004MNRAS.353..624D}. Convergence in phase-space structure is weaker, for the first time, numerous dark matter streams could be found within the VL2 simulation.

%Similarly, one finds that the mass or 
%circular velocity function of subhaloes has converged in the outer halo, but increasing the 
%resolution leads to more substructures in the inner regions where tidal forces can dissolve 
%poorly resolved structures in lower resolution simulations. 
%When one achieves a resolution 
%of several hundred million particles within the virial radius a wealth of structure in the 
%inner phase-space structure starts to become apparent, highlighting the fact that convergence 
%properties depend on the physical quantity of interest.

There are several reasons why we wish to carry out further studies at a higher resolution:
(i) There are many old and forthcoming observational tests that constrain the structure of 
dark matter haloes on scales well within $0.001R_{\rm vir}$. These include high resolution rotation curve data and the kinematics of stars at the centres of dwarf galaxies. Future proper motions of these inner stars with GAIA or SIM will provide even tighter constraints. The close binary nuclei in galaxies such as VCC128 
constrains the dynamics on even smaller scales \citep{2008arXiv0806.1951G}. 
(ii) As large surveys have pushed the surface brightness limits and detection efficiencies, many 
extremely faint satellite galaxies have been found orbiting the Milky Way. The completeness 
of current surveys is debated, and it has been argued that many hundreds of additional 
systems may be found in the coming years \citep{2008arXiv0806.4381T}. Simulations that can resolve 
and follow the survival of substructure within $10\%$ of $R_{\rm vir}$ are necessary to compare with these data.
(iii) Dark matter detection, either directly on Earth or indirectly via detection of annihilation 
relics, is the ultimate way to determine its nature. These experiments rely on accurate 
predictions for the phase-space structure of dark matter at the position of the Earth's 
orbit and the abundance and inner properties of substructure throughout the Galactic halo.
(iv) Understanding the equilibrium structure
resulting from violent relaxation is the ultimate challenge for galactic dynamicists. There is no compelling 
theory that can explain universal density and phase-space density profiles  \citep{2001ApJ...563..483T}, 
or correlations such as between the local density profile and the anisotropy parameter \citep{2006JCAP...05..014H}. 
%A successful theory should be able to explain these results from simulations, as well as 
%being able to predict the behaviour on smaller scales as $r\rightarrow 0$.

Given this motivation, we have carried out a sequence of simulations of a single Galactic mass dark matter halo, which at our highest resolution contains over a billion particles within its virial radius. In this letter we report on its inner structure and convergence properties.

%for the first time an analysis of the central density profile of a CDM halo to a radius of 
%$0.03\% of R_{\rm vir}$. We will compare with lower resolution simulations of GHALO and the VL2 halo 
%to measure the convergence of the density profile, substructure properties and phase-space distribution.

\begin{figure}
\label{pic}
\includegraphics[width=\hsize]{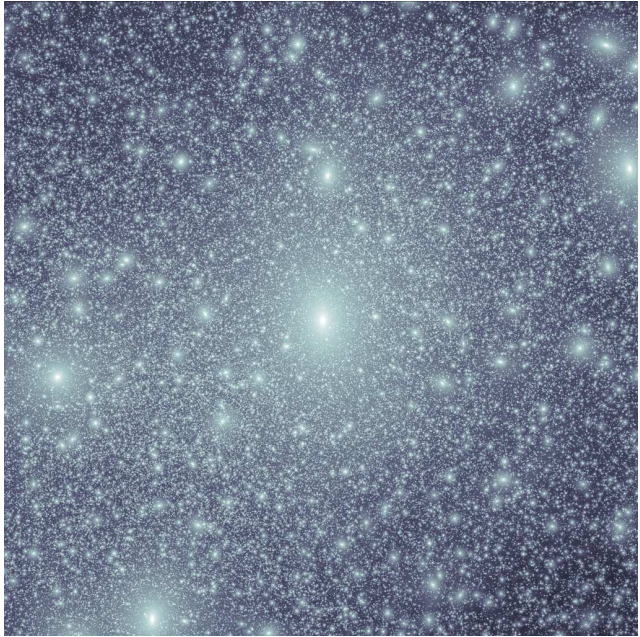}
\caption{The density of dark matter within the inner 200 kpc of GHALO$_2$. There are
about 100,000 subhaloes that orbit within the virial radius. Each bright spot in this image is an individual, bound, dark matter subhalo made up of many thousands of particles (there are far more particles than pixels here).}
\end{figure}

\section{The Simulations}

Our initial conditions are based upon the WMAP3+SDSS
\citep{2007ApJS..170..377S,2006JPhG...33....1Y} cosmological model with 
$\sigma_8 = 0.742, \Omega_{M}=0.237, \Omega_{\Lambda}=0.763, h=0.735, n=0.951$. 
The galaxy sized, $10^{12} M_{\odot}$, $R_{\rm vir} = 240$ kpc, 
halo was selected from a cosmological cube of 40 Mpc on a side. This simulation had $488^3$ particles
(simulation GHALO$_5$) in which three further nested spatial refinements by a factor of 3 
(GHALO$_{4, 3, 2}$) were placed such that the Lagrangian region of about $3 R_{\rm vir}$ of the 
halo at $z=0$ was covered by $2.1 \times 10^9$ high resolution particles in the initial condition. 
The final effective resolution of GHALO$_2$ is $13176^3$ resulting in a particle mass of 1000~$M_{\odot}$ and 
a total of $3.1 \times 10^9$ particles and $1.3 \times 10^9$ particles within $R_{200}=347$ kpc.
This allows us to capture all substructures out to more than $2 R_{200}$ at the highest 
resolution. A further refinement GHALO$_1$ (in progress) will resolve the phase-space structure 
at the position of the sun more sharply for future recoil dark matter detection experiments.

Creating these initial conditions was a significant challenge and we
had to parallelize the GRAFIC1 {\em and} GRAFIC2 codes of \cite{2001ApJS..137....1B}
whereby the GRAFIC2 code was completely rewritten in C and MPI, and checked for near machine 
precision agreement with the original GRAFIC2. The new parallel GRAFIC1\&2 codes can be obtained from the authors.
Generation of the initial condition 
%(once the regions of refinement had been determined) 
took 10 hours on 500 CPUs. 
We found that the original GRAFIC2 code had a bug in which the power spectrum used for the refinements 
was effectively that of the baryonic component. Although this has affected many previous simulations (not GHALO, nor VL2), 
tests show that the conclusions of these studies are not compromised.
%GHALO has used the CDM power spectrum for all refinements.

The GHALO$_2$ simulation was run at the Barcelona Supercomputer Center on 1000 CPUs of 
Marenostrum using a total of 2 million CPU hours. Several significant improvements 
to the gravity code PKDGRAV2 made this calculation possible including much better parallel 
computing efficiency and SIMD vector processing.
PKDGRAV2 uses a fast multipole method (FMM) similar to \citet{2000ApJ...536L..39D,2002JCoPh.179...27D} but using a 
5th-order reduced expansion for faster and more accurate force calculation in parallel, and a 
multipole based Ewald summation technique for periodic boundary conditions 
\citep{2001PhDT........21S}. 
It uses adaptive individual time-steps for particles based on a new estimator of the local dynamical time \citep{2007MNRAS.376..273Z}. 
%The usual criterion for selecting a time-step for a particle, $i$, in a cosmological simulation has been 
%$\Delta t_i = \eta \sqrt{\epsilon_i / |{\bf a}_i|}$, where $\epsilon$ is the gravitational 
%softening of the particle, ${\bf a}$ its acceleration, and $\eta$ a constant typically set 
%in the range 0.2--0.3. As this criterion is not related to the dynamical time in the 
%simulation it either takes too few steps in the very inner region of the halo and/or 
%takes too many steps in the outer regions of the halo. For this reason we use an efficient 
%variant of the dynamical time-stepping method of \cite{2007MNRAS.376..273Z} that can be used in an FMM code.
The opening angle in the gravity tree and the accuracy parameter in the dynamical time-stepping is 
$\Theta = 0.55$ and $\eta = 0.03$ before $z=2$, and then increased to 0.7 and 0.06 respectively.
We make several comparisons to the VL2 simulation which was also run with the FMM version
of PKDGRAV2, but whose initial conditions were selected and generated independently using 
somewhat different methods. 
%Much better parallel computing efficiency and better use of 
%SIMD vector processing allowed GHALO$_2$ (4 times the floating point operations) to 
%complete using about the same resources as VL2.
The VL2 halo has a mass of $2 \times 10^{12} M_{\odot}$ and used a particle mass of 4000 $M_{\odot}$.
The spline softening lengths for GHALO$_2$, VL2, GHALO$_{3, 4, 5}$ are 61, 40, 182, 546, and 1639 pc, 
respectively (for GHALO these are set to 1/50 of the mean inter-particle separation).

\section{The inner halo structure}

\subsection{The dark matter density profile}

\begin{table*}
\begin{tabular}{lc|cc|cc|cc|cc}
 & & \multicolumn{2}{c}{$\Delta^2 (\times 10^{-4})$} & \multicolumn{2}{l}{$\rho\:[10^6 M_{\odot}kpc^{-3}]$} & 
\multicolumn{2}{c}{$R\;[kpc]$} & \multicolumn{2}{c}{3rd parameter} \\
Fitting Function & Hernquist $(\alpha,\beta,\gamma)$ ($\rho_s,R_s$) & GH$_2$ & VL2 & GH$_2$ & VL2 & GH$_2$ & VL2 & GH$_2$ & VL2 \\
\hline
NFW & (1,3,1) & 3.5 & 6.6 & 2.32 & 4.24 & 14.1 & 13.9 & \multicolumn{2}{c}{-----} \\
Dehnen-McLaughlin & $(\slantfrac{4}{9},\slantfrac{31}{9},\slantfrac{7}{9})$ & 
1.6 & 0.70 & 0.273 & 0.591 & 42.6 & 36.7 & \multicolumn{2}{c}{-----}\\
S\&M-profile ($\rho_0,R_{\lambda}$) & ----- & 
{\bf 0.93} & {\bf 0.41} & 5050 & 11000 & 2.20 & 1.88 & \multicolumn{2}{c}{-----}\\
Generalized NFW & (1,3,$\gamma$) & 3.0 & 2.7 & 1.78 & 1.87 & 16.2 & 20.9 & 1.04 & 1.13 \\
Dehnen-McLaughlin & $(\slantfrac{(4-2\beta_0)}{9},\slantfrac{(31-2\beta_0)}{9},\slantfrac{(7+10\beta_0)}{9})$ & 
1.3 & 0.68 & 0.466 & 0.522 & 32.0 & 39.1 & -0.0531 & 0.0129 \\
Prugniel-Simien ($\rho',R_e,\alpha$)& ----- & 1.5 & 0.94 & 14.0 & 19.5 & 59.6 & 92.4 & 0.376 & 0.328 \\
Einasto ($\rho_{-2},R_{-2},\alpha$)& ----- & 1.0 & 0.45 & 0.685 & 0.991 & 26.8 & 28.9 & 0.155 & 0.142 \\
S\&M-profile ($\rho_0,R_{\lambda},\lambda$) & ----- & {\bf 0.92} & {\bf 0.41} & 4710 & 11200 & 2.47 & 1.82 & {\bf 0.102} & {\bf 0.100} \\
%\hline
%GHALO$_2$ &  & c-axis & a-axis & c-axis & a-axis & c-axis & a-axis & c-axis & a-axis \\
%\hline
%Einasto (Axial) & ----- & 1.1 & 1.9 & 0.649 & 1.25 & 23.8 & 25.0 & 0.152 & 0.177 \\
\end{tabular}
\caption{Fitting parameters and $\Delta^2$ for each of the 2 and 3-parameter models for both GHALO$_2$ and VL2 simulations. 
Here $\Delta^2 = \sum^m_i (\ln(\rho_i)-\ln(\rho_{\rm MODEL}(r_i)))^2/(m-3)$ where $\rho_i$ are the density values in  
logarithmically spaced radial bins at $r_i$. We fit from the resolved radius to 15\% of $R_{\rm vir}$ at which point 
substructure begins to cause significant fluctuations in the profile. Consistent with \citep{2008arXiv0805.1244D} we obtain 
a generalized NFW with ($\rho_s,R_s,\gamma$) = (1.05,28.0,1.23) (units as above) for VL2 by fitting from 360 pc to 
$R_{\rm vir}$, with the best fit profile being Prugniel-Simien over this range, $(\rho',R_e,\alpha) = (18.3,113,0.308)$.}
\end{table*}

We apply a logarithmic binning to determine the radial density profile for the various 
simulations which are shown in Figure~2. The convergence radius of the density profile for 
the lower resolution realizations (GHALO$_{3, 4, 5}$) can be clearly seen and are shown by the tick marks.
These scale roughly as expected with $r_{\rm conv} \propto N^{-1/3}$, 
%although the convergence radius of GHALO$_5$ is somewhat worse than this. 
and we extrapolate this to conclude that the 
convergence radius of GHALO$_2$ is around 120 pc. The inner slope of GHALO$_2$
is -0.8 at 120 pc = $0.05\%$ of $R_{\rm vir}$ and -1.4 at 2 kpc where the first subhalos become visible. Also 
shown is the power-law slope as a function of $\log(r)$, which exhibits a similar linear functional form for both haloes with no rescaling. 
Based on this observation we propose a new functional form for the fitting function of the density profile,
\begin{equation}
\label{eqn:lambda}
\rho(r) = \rho_0 e^{-\lambda(\ln(1 + r/R_{\lambda}))^2}
\end{equation}
which we term the S\&M-profile (Stadel \& Moore in preparation). It is linear in this plot down to a scale $R_{\lambda}$ beyond 
which it approaches the central maximum density $\rho_0$ as $r \rightarrow 0$. We also note that if 
one makes a plot of $d\ln\rho/d\ln(1+r/R_{\lambda})$ vs. $\ln(1+r/R_{\lambda})$, then this profile
forms an exact straight line with slope $-2\lambda$.

\begin{figure}
\label{dprof}
\includegraphics[width=\hsize]{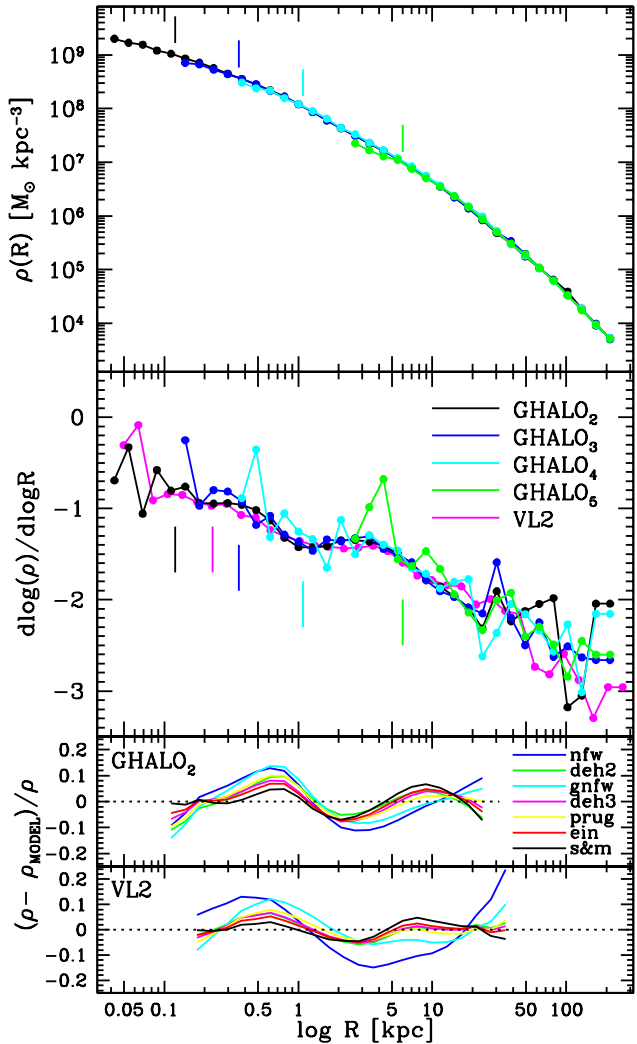}
\caption{The upper panel shows the density profile of GHALO$_2$ and its lower resolution realizations as well as
the density profile of the VL2 simulation in magenta. The convergence radius at each step in resolution
is easily seen (indicated by the tick marks). 
%showing that the convergence radius does scale roughly as 
%$r_{\rm conv} \propto N_{\rm vir}^{-1/3}$. The middle panel shows the logarithmic slope for each of the simulations. 
The lower panel shows the residuals of the GHALO$_2$ simulation with respect to 2-parameter fitting functions:
NFW (blue) and Dehnen-McLaughlin (green); as well as 3-parameter fitting functions: S\&M-profile (black), 
Einasto (red), Generalized NFW (cyan), Dehnen-McLaughlin (magenta), Prugniel-Simien (yellow).}
\end{figure}

Table~1 lists the best fitting parameters for several functions: the S\&M-profile,
 the restricted Hernquist $(\alpha,\beta,\gamma)$ 
profiles \citep{1990ApJ...356..359H,1996MNRAS.278..488Z}, the Einasto profile \citep{1969Afz.....5..137E,2004MNRAS.349.1039N}
\begin{equation}
\label{eqn:einasto}
\rho(r) = \rho_{-2}\exp(-\slantfrac{2}{\alpha}[(r/R_{-2})^\alpha-1]),
\end{equation}
and the \citet{1997A&A...321..111P} profile
\begin{equation}
\label{eqn:prug}
\rho(r) = \rho'(r/R_e)^{-p_\alpha}\exp(-b_\alpha(r/R_{-2})^\alpha),
\end{equation}
where $p_\alpha = 1-0.6097\alpha+0.05463\alpha^2$ and $b_\alpha = \slantfrac{2}{\alpha}-\slantfrac{1}{3} + 0.009876\alpha$ 
(for $\alpha < 2$, see \citet{2006AJ....132.2685M}) such that when projected one obtains a S\'ersic profile \citep{1963BAAA....6...41S,1968adga.book.....S}.

The residuals shown in Figure~2 show that the S\&M-profile provides a slightly better fit than all the 
models for the inner, more consistent, part of the profile. Furthermore, it is the only 3-parameter model where the 3rd 
parameter has a consistent value for the two different simulations. For this reason we also list this model as a possible 
2-parameter model, fixing $\lambda=0.1$. The Einasto profile also provides an excellent fit to the density profiles of 
the two simulations. 
%If we extend the fit to the virial radius, including the region showing a high degree of scatter 
%in the logarithmic slope, then the fits are less consistent and the best fit for GHALO$_2$ is then the Einasto profile 
%while for VL2 it is the Prugniel-Simien profile.

%Table 1 summarizes the fits to these functions, as well as the $\lambda$-profile given by equation~1 and
%the residuals over the range of the fits are shown in the bottom 2 panels of figure~2. 
%With just two simulations
%probing the inner parts of the density profile it is difficult to really favour one 3-parameter profile over %another, although all useful functions should show the same nearly linear dependence of $d\log\rho/d\log r$ with $\log r$.
% should we write more here about the fits, which is stated in the table caption? We could also say more in the 
% conclusions about these fits.

\subsection{Convergence of Halo Shape}

The convergence of the shape parameters (see also \citet{2006MNRAS.367.1781A}) for GHALO in Figure~3 show that it is highly prolate over all resolved regions with b/a = c/a $\approx$ 0.5. At the halo centres the shape diverges quickly to a more spherical configuration. This is likely due to the orbital distribution being modified by the effects of resolution and softening. In this region the velocity distribution function is also strongly affected. 

We estimate the convergence in the shape to be achieved at 0.3, 0.6, 2, 15 kpc for GHALO$_{2,3,4,5}$ respectively, a radius that is about 3 times the inferred convergence radius of the density profile but also scaling as
$N^{-1/3}$. 
The fact that the variation in shape has little impact on the density profile can be understood by comparing the density 
profile taken in a 15 degree cone about the major, $a$, axis and the minor, $c$, axis \citep{2002ApJ...574..538J}. The $\Delta^2$ for the fits to the 
various density profiles remains roughly consistent between the two axial density profiles, although
the best fit parameters vary. Due to the prolate shape the density profile parameters for the short axis are similar to the ones presented in Table~1. 
%
% do we want to say anything more profound about the shape?
%

\begin{figure}
\label{shape}
\includegraphics[width=\hsize]{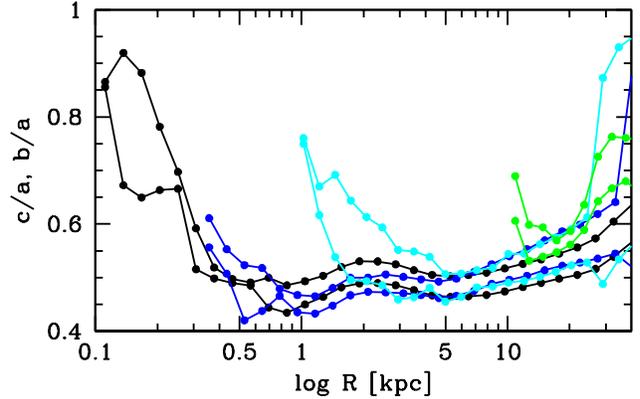}
\caption{Shape parameters for GHALO$_{2,3,4,5}$.}
\end{figure}

\subsection{Phase-Space Density Profile}

It has been pointed out \citep{2001ApJ...563..483T,2005MNRAS.360..892D,2005MNRAS.363.1057D} 
that the phase-space density (PSD) proxy, $\rho\sigma^{-3}$ vs $R$ is a power-law for CDM haloes,
and several new fitting functions for the density profile have been proposed using 
this fact as a starting point such as the Dehnen-McLaughlin models. When averaged in shells, 
$\rho(2\pi\sigma^2)^{-3/2}$ is remarkably well fit by a power-law with slope of $-1.84$ as shown in Figure~4.
However, it is interesting to compare this spherically averaged estimate with the true 6-dimensional PSD.

The code EnBiD \citep{2006MNRAS.373.1293S} has improved on earlier work by \citet{2005MNRAS.356..872A} in calculating better 
estimates of the 6-dimensional phase-space volume occupied by each particle and hence the PSD.
Taking the mean EnBiD PSD in logarithmic shells we see that the closest subhalo at 1.8 kpc stands out 
prominently and subhalos at larger radii begin to dominate the mean. 
Using a method based on a 6-dimensional Voronoi tessellation \citet{2004MNRAS.353...15A} also showed that the subhalos form a 
dominant contribution to the phase-space density.
This feature of using the
EnBiD PSD can be turned to great advantage in identifying subhalos and other substructures such as phase-space streams. 
However, removing the effect of subhalos with $f_{\rm EnBiD} > 100\;M_{\odot}{\rm kpc}^{-3}{\rm (km/s)}^{-3}$ from the mean,
we extend the mean background PSD out to much larger radii as shown in Figure~4. 
By removing streams, with $f_{\rm EnBiD} > 0.4 M_{\odot}{\rm kpc}^{-3}{\rm (km/s)}^{-3}$, we can 
extend this to at least 40 kpc.

We find that the true radial PSD profile estimated with EnBiD does not follow such a perfect power law and shows a 
steeper slope (roughly $-2$) than the $\rho\sigma^{-3}$ estimator. 
The EnBiD mean estimate and $\rho(2\pi\sigma^2)^{-3/2}$ are in agreement from about 0.2 to 2 kpc but the meaning 
of the power-law behaviour of $\rho(2\pi\sigma^2)^{-3/2}$ is unclear given that inside of 0.2 kpc it is under-resolved 
and outside of 2 kpc a large contribution comes from the substructure. A further concern is the considerable variation of 
$\rho\sigma^{-3}$ about a spherical shell of the prolate inner halo, which
makes it remarkable that we obtain the same power-law slope as originally found by 
\citep{2001ApJ...563..483T} despite the averaging that is taking place. This also explains the good
performance of the Dehnen-McLaughlin 2 and 3-parameter models at fitting the density profile.

From about 2 to 40 kpc the $\rho\sigma^{-3}$ estimator
is somewhat enhanced due to the presence of substructure, while inside of 0.1 kpc the EnBiD--mean continues to
resolve the power-law behaviour of the profile.

%The origin of the power-law PSD distribution function 
%$v(f)$ \citep{2004MNRAS.353...15A} is currently not understood, nor is it know whether there exists a 
%clear connection between this and the radial power-law observed in the PSD profile. 

\begin{figure}
\label{plot4}
\begin{center}
\includegraphics[width=\hsize]{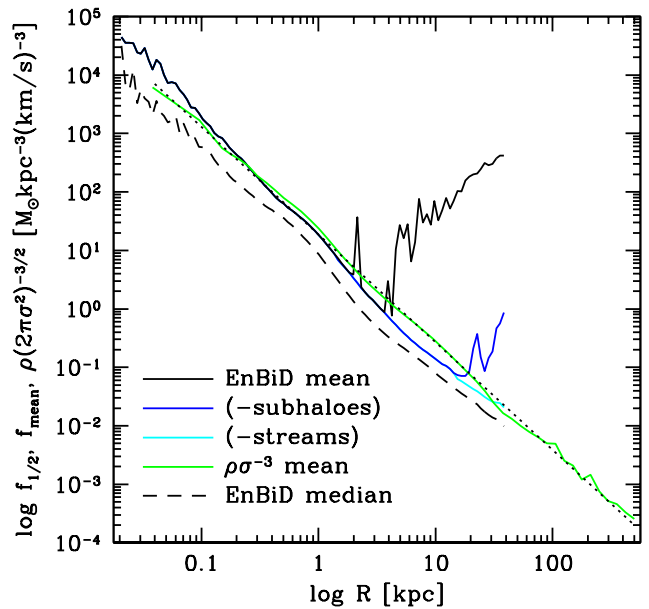}
\end{center}
\caption{The phase-space density profile of the main halo, measured in several different ways is shown.
The solid green line shows the traditional $\rho(2\pi\sigma^2)^{-3/2}$ averaged in shells. The solid black
and dashed black curves shows the mean and median EnBiD phase-space density estimator \citep{2006MNRAS.373.1293S} 
for the particles in logarithmic shells extending out to 40 kpc. The blue and cyan curves show the mean 
EnBiD phase-space density profile, but where the contribution from subhalos (blue) and subhalos+streams (cyan) 
has been excluded. Despite the effects of substructure the $\rho\sigma^{-3}$ profile is remarkably well fit by 
a power-law with slope $-1.84$ shown as the dotted black line. 
}
\end{figure}

\section{Conclusions}

The GHALO$_2$ simulation has achieved an unprecedented spatial and mass resolution within a CDM halo,
resolving thousands of subhalos within a radius corresponding to the 
galactic disk and a rich phase-space structure of streams beyond a radius of $\sim 8$ kpc.
Whilst there are more detailed analyses of this simulation in progress, we have reported here 
on the global inner properties of density and phase-space density profiles and halo shape.
Using a sequence of simulations of the same halo at difference resolutions, from $10^5$ -- $10^9$ particles,  
we confirm that the convergence radii for the density profile and shape scales as $N_{\rm vir}^{-1/3}$. 
The logarithmic slope of the radial density profile is close to a power law, gradually turning over to a slope 
of $-0.8$ at our innermost resolved region ($0.05\%$ of $R_{\rm vir}$). We have proposed a new two parameter fitting 
function that has a linearly varying logarithmic gradient which provides the best fit to the inner part of the 
GHALO and VL2 haloes. 
A larger sample of haloes, such as \citet{2001MNRAS.321..559B} and \citet{2007MNRAS.378...55M}, would be required to determine if this 
functional form provides a universal fit. We find that the convergence radius of the density is a factor of three 
smaller than the convergence of halo shape. GHALO is prolate, yet becomes spherical within a region where orbits are 
most likely innacurately followed due to the effects of finite particle number, relaxation and softening.

All functional forms fit to density profiles, whether 2 or 3 parameters are
empirical fits, even those based on properties (the Dehnen-McLaughlin) of the phase-space density profile 
whose origin is still poorly understood. Therefore the only current confidence can be given
to those profiles which have been fit to the highest resolution simulations and over the widest range of halos
encountered in N-body simulations. Clearly these two criteria are in conflict since simulations at the resolution 
of GHALO are too expensive to allow a broader study. Therefore, the results presented here should be considered as 
guides only, whose generality remains to be tested. Never the less we can consider economy of parameters and 
simplicity of functional form as guiding principles in the search for suitable profile functions to describe the 
end state of gravitational collapse. All the profiles we fit here (Table~1 and residuals in Figure~2) 
meet these subjective criteria, having at most 3 free parameters and simple functional forms.

While the phase-space density estimated by $\rho\sigma^{-3}$ is observed to follow a power 
law in radius of slope $-1.84$, its meaning is less clear since at small radii it is limited by
resolution of the estimator and at larger radii it becomes dominated by subhalos. Using the more
sophisticated EnBiD PSD estimator we find that the radial profile is steeper with an index of 
about $-2$, but that it is not as perfect a power-law as seen in $\rho\sigma^{-3}(r)$.

As a final comment, we note that in large galaxies, the inner structure and shape of the dark matter halo has 
likely been altered over time by the baryons via a range of physical effects, including dissipation, energy 
transfer from sinking massive objects, binary black holes, bar-halo interactions, 
turbulent gas motions and more. Simulations that follow the baryonic components together 
with the dark matter will resolve these additional questions in the coming years. 

%For smaller galaxies, 
%the density of baryons today is insufficient to change the properties of their dark matter 
%haloes. Thus besides addressing the theoretical question of violent relaxation and virialisation, 
%we can use dark matter simulations to address challenges from observational data.

\section*{Acknowledgments}

We are grateful to the Barcelona Supercomputing Center for their
generous allocation of resources on the Marenostrum supercomputer and 
to Jose Maria Iba\~nez for his enthusiastic support of this project. 
Support for this work was 
also provided by the Spanish Ministry grant MEC AYA2007-67752-C03-02 (V.Q.)
and by NASA through grants HST-AR-11268.01-A1 and NNX08AV68G (P.M.) 
and Hubble Fellowship grant HST-HF-01194.01 (J.D.).

%\bibliography{ghalo.bib}
%
% For astro-ph comment out the above and uncomment the line below *after* you 
% have successfully built the paper, i.e., have a valid ghalo.bbl file!
%
\bibliography{ghalo.bbl}

\label{lastpage}

\end{document}